\documentclass[showkeys,showpacs]{revtex4}
\usepackage{amsmath}
\usepackage{amssymb}
\usepackage{graphicx}

\begin{document}

\title{Yamabe flow, conformal gravity and spacetime foam}
\author{Vladimir Dzhunushaliev
\footnote{Senior Associate of the Abdus Salam ICTP}}
\email{vdzhunus@krsu.edu.kg}
\affiliation{
%ASC, 
%Department f{\" u}r Physik, Ludwig-Maximilians-Universit{\" a}t M{\" u}nchen,
%Theresienstr. 37, D-80333, Munich, Germany \\
%and \\
Dept. Phys. and Microel. Engineer., Kyrgyz-Russian Slavic University, Bishkek, Kievskaya Str.
44, 720021, Kyrgyz Republic}

\begin{abstract}
It is shown that 3D part of a spherically symmetric solution in conformal Weyl gravity interacting with Maxwell electrodynamics is a Yamabe flow as well. The Yamabe flow describes the transition from a horn of an initial wormhole to a 3D Euclidean space both filled with a radial electric field. It is supposed that such transition may describe appearing/disappearing quantum wormholes in a spacetime foam. 
%
%2000 MSC: 53C44, 53C21, 83D05, 83E99
\end{abstract}

\maketitle

\section{Introduction}

In differential geometry, the Ricci flow is an intrinsic geometric flow: a process which deforms the metric of a Riemannian manifold. Ricci flow is an analog of the heat equation for geometry, a diffusive process  acting on the metric of a Riemannian manifold, in a manner formally analogous to the diffusion of heat, thereby smoothing out irregularities in the metric. The Ricci flows wese introduced by Hamilton \cite{Hamilton} over 25 years ago. It plays an important role in the proof of the Poincare conjecture \cite{Perelman}. Obviously, one could introduce many different flows. In this paper we work with a Yamabe flow. The physical application of Ricci flow can be found in Ref's~\cite{Headrick:2006ti} - \cite{Vacaru:2007nq}. 
\par 
In 1918 Hermann Weyl proposed \cite{weyl} a new kind of geometry and a unified theory of gravitation and electromagnetism based on it. Fourth order metric theories of gravitation have been discussed from 1918 up to now. One original motivation was the scale invariance of the action, a property which does not hold in general relativity. Another motivation was the search for a unification of gravity with electromagnetism, which is only partially achieved with the Einstein-Maxwell system. In 1966 a renewed interest in these theories arose in connection with a semiclassical description of quantum gravity \cite{Pechlaner}-\cite{Treder}. The fact that fourth order gravity is one-loop renormalizable in contrast to general relativity was \cite{Stelle} initiated a boom of research. Also the superstring theory gives in the field theoretical limit a curvature-squared contribution to the action \cite{Ivanov} \cite{Hochberg}. In Ref. \cite{bach} Rudolf Bach realized the possibility to have the conformal invariance in a purely metrical theory. The history of fourth order metric theories of gravitation can be found in Ref.~\cite{Schimming:2004yx}.
\par
Spacetime foam is a concept in quantum gravity, devised by John Wheeler in 1955 \cite{Wheeler}.
It is postulated that the spacetime foam is a cloud of quantum wormholes with a typical linear size of the Planck length \cite{Hawking}. Schematically in some rough approximation we can imagine the appearing/disappearing of quantum wormholes as emergence, growth and meshing of two horns into a wormhole with a subsequent rupture of the quantum wormhole. For the macroscopic observer these quantum fluctuations are smoothed and we have an ordinary smooth manifold with the metric submitting to Einstein equations. The spacetime foam is a qualitative description of the turbulence that the phenomenon creates at extremely small distances of the order of the Planck length. 
The appearance/disappearing of these quantum wormholes leads to the change of spacetime topology. This fact give rise to big difficulties at the description of the spacetime foam as by the topology changes of a space (according to Morse theory) the critical points must exist where the time direction is not defined. In each such point should be a singularity which is an obstacle for the mathematical description of the spacetime foam.
\par
In this Letter we show that there exists a connection between above mentioned fields of mathematics and physics. We show that the 3D part of a spherically symmetric solution in a conformal gravity + Maxwell electrodynamics is simultaneously a Yamabe flow and describes the transition from one horn of a wormhole to a flat space filled with a radial electric field. 

\section{Yamabe flow and spherically symmetric solution in conformal gravity}

The Yamabe flow is defined by the equation \cite{hamilton2}
\begin{equation}
	\partial_{\tau}g_{ij} = -R g_{ij}
\label{1-10}
\end{equation}
where $g_{ij}, i,j = 1,2,3$ is a metric, $R$ is a 3D scalar curvature. Under this flow, the conformal class of a metric does not change.
\par 
The lagrangian for the conformal Weyl gravity interacting with the U(1) gauge field is: 
\begin{equation}
	{\cal L} = -C_{\mu \nu \rho  \sigma} C^{\mu \nu \rho  \sigma} - 
	\frac{\kappa}{4} \mathrm{tr}
	\left (F_{\mu \nu}F^{\mu \nu} \right ) ,
\label{1-20}
\end{equation}
here $C_{\mu \nu \rho  \sigma}$ is the conformally invariant Weyl tensor, $F_{\mu \nu}$ is the tensor of field  strength  for the electromagnetic potential, $\mu, \nu = 0,1,2,3$. The Bach equations for Lagrangian \eqref{1-20} are: 
\begin{equation}
	B_{\mu \nu} = \kappa T_{\mu \nu} ,
\label{1-30}
\end{equation}
where 
\begin{eqnarray}
	B_{\mu \nu} &=&  B^{(1)}_{\mu \nu}  + B^{(2)}_{\mu \nu} ,
\label{1-40}\\
	B^{(1)}_{\mu \nu} &=&
	- \Box R_{\mu \nu} + 2 R^{\alpha}_{\mu \ ;\nu \alpha} - \frac{2}{3} R_{;\mu \nu}
	 + \frac{1}{6} g_{\mu \nu} \Box R , 
\label{1-50}\\
	B^{(2)}_{\mu \nu}  &=& \frac{2}{3} R \ R_{\mu \nu}
	- 2 R_{\mu \alpha}R^\alpha_\nu - \frac{1}{6} R^2 g_{\mu \nu}
	+ \frac{1}{2} g_{\mu \nu} R_{\rho \sigma} R^{\rho \sigma} 
\label{1-60}
\end{eqnarray}
and $T_{\mu \nu}$ is the energy-momentum tensor for the electromagnetic field. 
\par 
In Ref.~\cite{riegert84} the static, spherically symmetric solution for the Bach-Maxwell equations is given: 
\begin{equation}
	ds^2 = \left ( \frac{r^2}{a_0} + br + c + \frac{d}{r} \right ) dt^2 - 
	\frac{dr^2}{\left ( \frac{r^2}{a_0} + br + c + \frac{d}{r} \right )}-
	r^2 d\Omega ^2 ,
\label{1-70}
\end{equation}
with an electromagnetic  one-form potential:
\begin{equation}
	A = A_i dx^i = A_t(r) dt ,
\label{1-80}
\end{equation}
where $a_0,b,c,d$ and $q$ are some constants with the following relation \eqref{1-90} where $q$ is the electric charge and the parameter $\beta$ used in \cite{riegert84} is related to our conventions via $ \beta \kappa = 4$: 
\begin{equation}
	3bd - c^2 + 1 + \frac{3}{8}q^2 \kappa = 0
\label{1-90}
\end{equation}
If $b=c=d=0$ we have the following solution: 
\begin{equation}
	ds^2 = r^2\left ( \frac{dt^2}{a_0} - \frac{a_0}{r^4}dr^2 - 
	d\Omega ^2\right ) .
\label{1-100}
\end{equation}
We can introduce  new dimensionless coordinates $t' = t/a_0$ and $x=\sqrt {a_0}/r$ then: 
\begin{equation}
	ds^2 = \frac{a_0}{x^2} \left ( dt^2 - dx^2 - d\Omega ^2\right )
\label{1-110}
\end{equation}
with $x\in(-\infty,+\infty)$ and we rename $t' \rightarrow t$. This metric is conformally equivalent to 
\begin{equation}
	ds^2 = dt^2 - dx^2 - d\Omega ^2 
\label{1-120}
\end{equation}
which represents the cartesian product of a flat and a non-flat 2-space of constant curvature. 
This solution is a tube filled by an electric field $E_x = F_{tx} = q, \quad F_{t \theta}=F_{t \phi}=0$.

\section{Yamabe transition from a wormhole to a flat space}

For the description of transition from a wormhole to a flat space we will use the 3D part of the metric 
\begin{equation}
	ds^2 = e^{2 \alpha(x, \tau)} \left( dt^2 - dx^2 - d\Omega ^2 \right)
\label{2-10}
\end{equation}
which is conformally equivalent to the metric \eqref{1-120} and $\tau$ is a Yamabe parameter controlling the Yamabe flow in Eq. \eqref{1-10}. Thus we use spacelike metric 
\begin{equation}
	dl^2 = e^{2 \alpha(x, \tau)} \left( dx^2 + d\Omega ^2 \right). 
\label{2-20}
\end{equation}
The substitution this metric into Yamabe equation \eqref{1-10} gives us 
\begin{equation}
	\dot \alpha = e^{-2 \alpha} \left(
		2 \alpha'' + {\alpha'}^2 - 1
	\right)
\label{2-30}
\end{equation}
here $\dot \alpha = \partial_\tau \alpha$ and $\alpha' = \partial_x \alpha$. At first we will search Yamabe solitons. 

\subsection{$\alpha_{1,2}$ Yamabe solitons}

Analogously to Ricci solitons the Yamabe solitons satisfy the condition 
$\dot \alpha = 0$. The equation 
\begin{equation}
		2 \alpha'' + {\alpha'}^2 - 1 = 0
\label{2-40}
\end{equation}
has two solutions 
\begin{eqnarray}
		\alpha_1(x) &=& x,
\label{2-50}\\
		\alpha_2(x) &=& \ln \left[ 4 \cosh^2 \left( \frac{x}{2} \right) \right].
\label{2-60}
\end{eqnarray}
At first we show that the first Yamabe soliton \eqref{2-50} describes a flat space. The metric is 
\begin{equation}
		dl^2 = e^{2x} \left( dx^2 + d \Omega^2 \right). 
\label{2-70}
\end{equation}
Let us introduce a new coordinate $r = e^x$ then the metric \eqref{2-70} is 
\begin{equation}
		dl^2 = dr^2 + r^2 d \Omega^2 
\label{2-80}
\end{equation}
which is the metric of the 3D Euclidean space ($\alpha_1$--flat Yamabe soliton). The electric field is 
$E_r = \frac{q}{r}$. 
\par 
Now we consider the second Yamabe solition \eqref{2-60}. The metric is 
\begin{equation}
		dl^2 = \left[ 4 \cosh^2 \left( \frac{x}{2} \right) \right]^2  
		\left( dx^2 + d \Omega^2 \right). 
\label{2-90}
\end{equation}
Let us introduce a new coordinate $dr = 4 \cosh^2 \left( \frac{x}{2} \right) dx$,  
$r = 2 \left( x + \sinh x \right)$ then the metric \eqref{2-90} is 
\begin{equation}
		dl^2 = dr^2 + a(r) d \Omega^2 
\label{2-100}
\end{equation}
where $a(r)$ is an even function and defined parametrically as follows
\begin{eqnarray}
		a(x) &=& \left[ 4 \cosh^2 \left( \frac{x}{2} \right) \right]^2,
\label{2-110}\\
		r(x) &=& 2 \left( x + \sinh x \right).
\label{2-120}
\end{eqnarray}
Asymptotically we have $a(r) \approx r^2$. It means that the second Yamabe soliton ($\alpha_2$--wormhole Yamabe soliton) describes an asymptotically flat wormhole.  The electric field is 
$E_r = \frac{q}{\sqrt{a(r)}}$. 
\par 
Thus the Riegert solution \eqref{2-10} describes both asymptotically flat wormhole and flat 3D space. In the next section we will show that there exists a Yamabe flow describing transition from nonasymptotically flat wormhole to Euclidean space and simultaneously the flow is the Riegert solution. 

\subsection{Yamabe transition}

\begin{figure}[h]
\begin{center}
\fbox{
 	\includegraphics{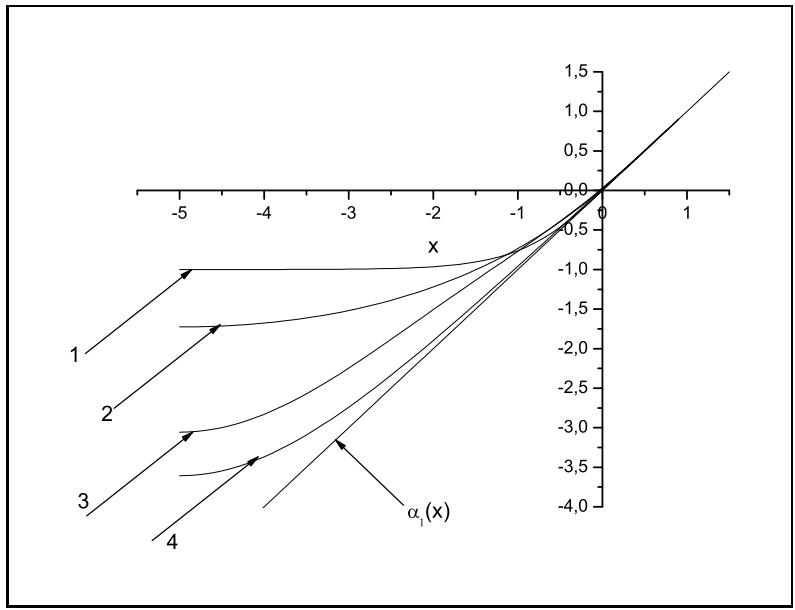}
}
	\caption{The profiles of the function $\alpha(x, \tau)$ for different $\tau$. 
	$\alpha_1(x)$ is the flat Yamabe soliton. 1 -- is the initial state; 
	2,3,4 are the functions $\alpha(x, \tau)$ for 
	$\tau = 0.09 \times t_0, 0.15 \times t_0, t_0$ correspondingly.}
\label{fig1}
\end{center}
\end{figure}

In this subsection we would like to present a numerical solution of Yamabe flow equation \eqref{2-30}. We use the following initial \eqref{2-130} and boundary conditions \eqref{2-140} \eqref{2-150} for the numerical solution of Yamabe equation \eqref{2-30}
\begin{eqnarray}
	\alpha(x, \tau=0) &=& 
	\left\{
	\begin{array}{rl}
		x,			&	\text{ if } x>0 \\
		\tanh x	&	\text{ if } x\leq 0 
	\end{array}
	\right. 
\label{2-130}\\
	\alpha(x=x_0, \tau) &=& x_0 ,
\label{2-140}\\
	\left. \left[ \frac{\partial}{\partial x}\alpha(x, \tau) \right] \right|_{x=x_1} &=& 
	\frac{1}{\cosh^2 x_1} 
\label{2-150}
\end{eqnarray}
The initial condition \eqref{2-130} is chosen in such a way that 
$\alpha(x>0, \tau=0) = \alpha_1(x)$ and boundary conditions \eqref{2-140} \eqref{2-150} are adjusted with the initial condition \eqref{2-130}.  The result of the numerical calculation is presented in Fig.~\ref{fig1}. 
\par
\begin{figure}[h]
\begin{center}
\fbox{
	\includegraphics[width=5cm,height=8cm,angle=-90]{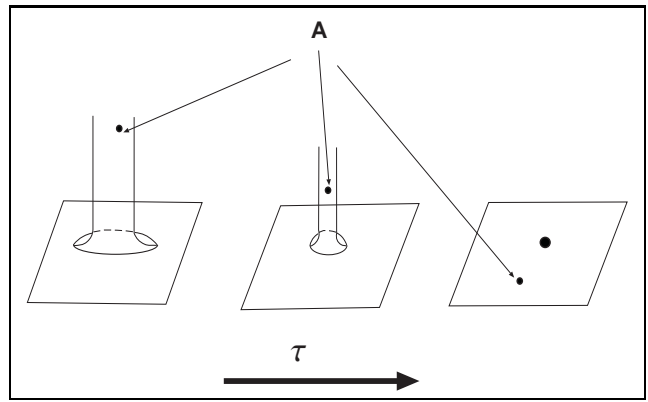}
}
	\caption{Schematical $\tau$ evolution of an initial horn to a final state. 
	\textbf{A} -- the position of the same point with increasing~$\tau$.}
\label{fig2}
\end{center}
\end{figure}

 
The initial condition \eqref{2-130} describes one horn of a wormhole: for $x>0$ the metric \eqref{2-20} is flat space but for $x \leq 0$ the metric describes a tube (horn). The result of numerical calculations shows that $\tau$ -- evolution of the initial condition is following: from the horn of initial wormhole (curve -- 1 in Fig.~\ref{fig1}) to the $\alpha_1(x)$ -- Yamabe soliton. Fig.~\ref{fig2} displays a $\tau$ evolution of the horn of initial wormhole to a final state (3D flat space).

\section{Discussion and conclusions}

We have found that 3D part of the Riegert solution in Weyl conformal gravity interacting with Maxwell electrodynamics is simultaneously the Yamabe flow. The properties of such flow are following:
\begin{itemize}
	\item The flow describes the disappearance of the horn of initial wormhole. 
	\item For each $\tau < \infty$ (auxiliary states) the length of the horn (the space with 
	$x < 0$) is infinite: $l_{x \in (-\infty , 0 ]} = \infty$. 
	\item For $\tau = \infty$ (final state) $l_{x \in (-\infty , 0 ]} < \infty$. 
%	\item The energy of electric field for any $\tau < \infty$ is infinite but for 
%	$\tau = \infty$ the energy is finite. 
	\item The corresponding 4D spacetimes are not asymptotically flat as 
	$g_{tt}(|x| \rightarrow \infty) \neq 1$. 
\end{itemize}
The consideration presented here allows us to say that the transition considered here may describe a quantum wormhole (handle) in a hypothesized spacetime foam. Also it can be considered as a indirect evidence that the Einstein gravity on the Planck level change to the conformal Weyl gravity. In this case the conformal Weyl gravity gives us the opportunity to avoid problems in Einstein gravity connected with a fundamental (Planck) length.

\section{Acknowledgments}
I am grateful to the Alexander von Humboldt Foundation for financial support and V. Mukhanov for invitation to Universit\"at M\"unich for research.

\end{document}